\documentclass[pra,aps,amsmath,amssymb,showkeys,showpacs,nofootinbib]{revtex4-2}
\usepackage{graphicx}
\newcommand{\xdif}{{\mathrm{d}}}

\newcommand{\rp}{{\mathtt{r}}}
\newcommand{\pr}{{\mathtt{P}}}
\newcommand{\mA}{{\mathsf{A}}}
\newcommand{\mB}{{\mathsf{B}}}
\newcommand{\mC}{{\mathsf{C}}}
\newcommand{\mD}{{\mathsf{D}}}
\newcommand{\ddt}{{\mathrm{det}}}
\newcommand{\itt}{{\mathtt{i}}}
\newcommand{\veps}{\varepsilon}
\begin{document}
\clearpage
\preprint{}

\title{A five-dimensional Lorenz-type model near the temperature of maximum density}

\author{Alexey E. Rastegin}
\email{alexrastegin@mail.ru}
\affiliation{Department of Theoretical Physics, Irkutsk State University, K. Marx St. 1, Irkutsk 664003, Russia}

\begin{abstract}
The current study formulates a convective model of the Lorenz type
near the temperature of maximum density. The existence of this temperature
actualizes water dynamics in temperate lakes. There is a conceptual
interest what this feature induces in Lorenz-type models. The
consideration starts with the zero coefficient of thermal expansion.
Other steps are like famous Tritton's approach to derive the Lorenz
model. This allows us to reduce difficulties with a selection of
Galerkin functions. The analysis focuses on changes induced by
zeroing the coefficient of thermal expansion. It results in a
five-dimensional Lorenz-type model, whose equations are all
nonlinear. The new model reiterates many features of the standard
Lorenz model. The nontrivial critical points appear, when the zero
critical point becomes unstable. The nontrivial critical points
correspond to two possible directions of fluid flow. Phase trajectories of the new model were studied numerically. The results are similar to the known five-dimensional extensions of the Lorenz model.
\end{abstract}


\maketitle

\pagenumbering{arabic}
\setcounter{page}{1}

\section{Introduction}\label{sec1}

Hydrodynamic instabilities and the transition to turbulence are
closely related problems with a long and wealthy history. The
discovery of strange attractors has provided a new approach to
understanding of the mechanisms by which turbulence may occur
\cite{berge,bhatta}. Many of the results obtained in this way were
stimulated by studies on the Lorenz model \cite{sparrow}. Edward
Lorenz laid the foundation for what many consider the third
revolution of 20th century --- chaos theory \cite{palmer}. The
Lorenz model \cite{lorenz63,essence} is recognized as a classic
realization of a simple physical system with strange attractors.
Following the ideas of Saltzman \cite{saltzman62}, it includes a
quite successful combination of properly chosen Galerkin modes. The
Lorenz equations used a very severe truncation of the equations of
B\'{e}nard convection. It is well known that a larger number of
Galerkin modes makes the model more adequate
\cite{curry,shen14,shen19,moon}. In this way, the whole family of
closely related models occurred. The dynamical system approach has
found its right place in the theory of turbulence \cite{bohr98} and
geophysical hydrodynamics \cite{tfja,rgwt}. It turns out that
low-order models are relevant to explore the scaling laws and
mechanisms of the energy cascade.

Pursuing various aims of a physical or even mathematical origin,
both simplified versions and generalizations of the standard Lorenz
model have been considered in the literature. From the point of view
of studying the structure of the Lorenz system itself, its
truncated versions were examined. In particular, the limiting
equations \cite{sparrow}, the geometric model \cite{gw79}, and the
nondissipative Lorenz model \cite{shen18}. Among the extensions of
the Lorenz system, models with a larger number of space modes in
the temperature function are best known \cite{shen14,shen19}. Models
with a larger selection of Galerkin modes in both the stream and
temperature functions are also known \cite{curry,moon,chilla}. Such models with an arbitrary number
of functions are rather the subject of numerical procedures.
However, in all of the listed cases, the thermal expansion
coefficient was assumed to be nonzero. In contrast, this study aimed
to consider a Lorenz-type model with zero coefficient of thermal
expansion. There are several reasons to accomplish this
consideration. Changes in the mathematical properties of the corresponding dynamical models are interesting themselves.
On the other hand, there is a
physical reason to deal with this question.

Substances common in everyday life typically undergo thermal
expansion in response to heating. In other words, a given mass of
a substance expands to a larger volume when its temperature increases.
Nevertheless, the coefficient of thermal expansion of some
substances can principally change the sign under certain conditions.
Water, especially important for Life, is also notable as having the
maximum density at temperatures near $4{\,}^{\circ}$C \cite{water10}.
Of course, the exact value of the temperature of maximum density of
natural water depends on actual pressure and salinity. In fact, the
annual evolution of stratification in temperate lakes according to
Forel's classification \cite{vb2014} is a shining
manifestations of the existence of this temperature. When
salinity effects are negligible, the temperatures of the upper layer of a
lake cross the point of maximum density twice in a year (see, e.g.,
subsection 1.3.1 of Ref. \cite{hutter}). The Great Lakes \cite{grady} are
an especially famous example. Lake Baikal known
for its outstanding features \cite{minoura} undergoes a
regular annual alternation of summer and inverse winter
stratification \cite{wck91}. There is a domain in the upper layer where
any linear relation between the density and temperature of water
becomes inadequate \cite{minoura}.

This study aims to examine changes in Lorenz-type systems near the
point of maximum density. The addressed question is interesting from
its own perspective and for applications in which the temperature of
maximum density plays an important role. Then, a buoyancy term
proportional to the square of excess temperature should be
considered. To investigate the question of interest, the coefficient
of thermal expansion was assumed to be zero in the setup. The
proposed model differs from all previous models by a nonlinear term
in the first equation, which succeeds the Navier--Stokes vector
equation. In effect, the obtained five-dimensional system contains
only nonlinear equations. To dive into the essence of the problem,
we treat it similarly to Tritton's physics-based approach to derive
the Lorenz model \cite{tritton}. This way allows us to mitigate
difficulties with selecting the required Galerkin functions. In a
more complete description, these functions satisfy the imposed
boundary conditions. This makes their selection more difficult than
in our adaptation of Tritton's approach.

This paper is organized as follows. Section \ref{sec2} describes the
physical basis of the model of interest. The accepted form of the
buoyancy term directly leads to a momentum equation with
nonlinearity in temperature deviations. In terms of properly chosen
dimensionless units, five differential equations with quadratic
nonlinearity appear in Sec. \ref{sec3}. The basic points of the
analysis of the proposed model are presented in Sec. \ref{sec4}. In
particular, we derive nontrivial critical points and discuss
conditions for their existence. As will be shown, the situation that
occurs in the standard Lorenz model is fully reproduced here. The
numerical investigation results of the proposed model are described
in Sec. \ref{sec5}. They allow us to support analytical findings and
analogies with the standard Lorenz model. In effect, projections on
different planes in phase space illustrate the existence of
attracting orbits. This situation is similar to that held for the
known five-dimensional extension of the standard model. Section
\ref{sec6} concludes the paper with a summary of the reported
results. The stability of nontrivial critical points is considered
in Appendix \ref{appa}.

\section{Physical basis of the model}\label{sec2}

Thermally induced convection is one of important directions of
researches in physics of fluids as well as in some engineering
disciplines. It is typically considered within the Boussinesq
approximation, different aspects of which are examined thoroughly in
Refs. \cite{veronis,mihal,dutton}. Rayleigh--B\'{e}nard convection is
studied in detail due to its analytical \cite{chandra,gershuni} and
experimental accessibility \cite{busse,chilla}. To study turbulence in
thermal convection, theoretical analysis \cite{kraichnan,martin} is
typically combined with experimental data
\cite{ahlers,xia02,sreeni02,sreeni05,xia08} and numerical simulation
\cite{crespo89,maxey89,silano,sreeni17,sreeni19}. Most of the
results were obtained for the case with the strictly positive
coefficient of thermal expansion. This coefficient reads as
\begin{equation}
\alpha=-\,\frac{1}{\rho}\biggl(\frac{\partial\rho}{\partial{T}}\biggr)_{\!{p}}
\, , \label{alp0}
\end{equation}
where $\rho$ is the density, $T$ is the temperature, and $p$ is the
pressure.

In contrast to typical situation with $\alpha>0$, the so-called
thermal contraction is also allowed in thermodynamics. In this
regard, the thermal expansion coefficient differs from the
coefficient of isothermal compressibility. The latter should be
positive due to classical thermodynamic (see, e.g., \S 21 of Ref.
\cite{landau5}). On the other hand, thermodynamic demonstrations
that compressibility must be positive all appear to deal with
additional assumptions \cite{lw2008}. These circumstances show that
convection with vanishing $\alpha$ deserves to be considered {\it a
fortiori}. Convective motions in water near the point of maximum
density have already found some coverage \cite{depaz,sonnino}. The
aim of the current study is to focus on other aspects. What happens
in models of the Lorenz type with zero $\alpha$? Thus, we will
further use a thermodynamic equation of state in the form
\begin{equation}
\rho=\rho_{0}\bigl[1-\varkappa(T-T_{0})^{2}\bigr]
\, . \label{rhokp}
\end{equation}
By $T_{0}$ and $\varkappa$, we mean here the temperature of maximum
density and the corresponding strictly positive factor. Variations
of height are assumed to be so small that their influence on the
temperature of maximum density is negligible.

\begin{figure}
\includegraphics[height=3.8cm]{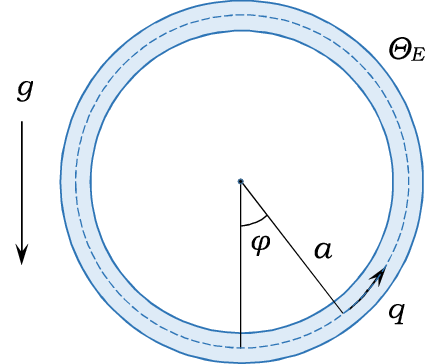}
\caption{\label{fig0} Fluid moving in the loop of radius $a$ under spatially heterogeneous temperature conditions.}
\end{figure}

The physical system of interest is shown in Fig. \ref{fig0}. The
fluid moves in a vertically placed loop of radius $a$ and fixed
cross section. In contrast to Tritton's approach to derive the
Lorenz system in section 17.3 of Ref. \cite{tritton}, the thermodynamic
equation of state reads as (\ref{rhokp}). Tritton's reformulation is
similar to the setup described in Ref. \cite{welander}. The position
round the loop is described by the angle $\varphi$. It is now
necessary to choose a profile of the external temperature. It often
occurs in physics that formulations with arbitrary functions on the
boundaries are difficult to tractable immediately. Then expansions
of some kind are typically used, for instance, the Taylor series
expansion. At least as the first step, the number of involved terms
is restricted to take into account nontrivial effects. It is helpful
to recall that the Lorenz system {\it per se} was built with minimal
number of Galerkin functions required to include a part of
nonlinearities of the initial equations. In this sense, the simplest
model of the whole family of ones is selected as the starting point.
It is well known that low-order models are physically sound in
various respects \cite{atpf99,tong}. A limited number of involved
equations can be enough to understand the scaling laws and
mechanisms of the energy cascade at small scales \cite{bohr98}.

According to the above reasons, the external temperature is taken to
vary with height according to
\begin{equation}
\varTheta_{E}=T_{0}+\varTheta_{1}\cos\varphi+\varTheta_{2}\cos2\varphi
\, , \label{thetex}
\end{equation}
where $\varTheta_{1}$ and $\varTheta_{2}$ are the prescribed
constants. In other words, the quantity $\varTheta_{E}$ is expanded
up to the term quadratic in the height deviation from the mean level
of the loop. We avoid to rewrite (\ref{thetex}) explicitly in terms
of the height deviation since only the values $\varTheta_{1}$ and
$\varTheta_{2}$ will be used in the following. It will be shown
that the quadratic term is required to make the consideration
self-consistent. Dealing with quadratic functions is naturally
suggested, when dependencies of the form (\ref{rhokp}) arise. The
chosen form restricted to a quadratic term is the first that leads
to nontrivial dynamics. In principle, more terms in the right side
of (\ref{thetex}) could be kept in future investigations.

It is assumed that the variations in velocity and temperature over a
cross section may be handled by working in terms of an average speed
$q(\varphi,t)$ and average temperature $T(\varphi,t)$. The speed $q$
is taken positive, when flow is anticlockwise. The continuity
equation corresponds to the Boussinesq approximation, viz.
\begin{equation}
\frac{\partial{q}}{\partial\varphi}=0
\, . \label{cont}
\end{equation}
Hence, $q$ depends on time only \cite{tritton}. The Navier--Stocks
equation reduces to the form that reflects the momentum balance.
Within the Tritton approach, it reads as
\begin{equation}
\frac{\xdif{q}}{\xdif{t}}=
-\!\;\frac{1}{\rho_{0}a}\>\frac{\partial{p}}{\partial\varphi}-\Gamma{q}-
\frac{\rho{g}\sin\varphi}{\rho_{0}}
\ . \label{rens}
\end{equation}
Following Ref. \cite{tritton}, the term with $\Gamma$ is placed in the
right-hand side of (\ref{rens}). Thereby, the action of viscosity is
supposed to induce a resistance to the motion proportional to its
speed. Like the Boussinesq approximation, density changes are taken
into account only in the third term in the right-hand side of
(\ref{rens}). It is valid since actual vertical acceleration is very
small in comparison with the gravitational one. Substituting
(\ref{rhokp}) into (\ref{rens}) and integrating with respect to
$\varphi$ results in
\begin{equation}
2\pi\,\frac{\xdif{q}}{\xdif{t}}=-\!\;2\pi\Gamma{q}+
g\varkappa\int_{0}^{2\pi}(T-T_{0})^{2}\sin\varphi\,\xdif\varphi
\, . \label{rens1}
\end{equation}
The last term in the right-hand side of (\ref{rens1}) replaces the
integral
\begin{equation}
g\alpha\int_{0}^{2\pi}(T-T_{0})\sin\varphi\,\xdif\varphi
\, , \label{tens1}
\end{equation}
which is used in Tritton's formulation. The term with the
coefficient of thermal expansion commonly appears within the
Boussinesq approximation. The current study aims to examine changes
due to zero $\alpha$. These changes are now reflected in the
right-hand side of (\ref{rens1}).

The fluid temperature inside the loop is represented in the form
\begin{equation}
T=T_{0}+T_{1}(t)\cos\varphi+T_{2}(t)\sin\varphi+T_{3}(t)\cos2\varphi
+T_{4}(t)\sin2\varphi
\, . \label{tempe}
\end{equation}
Tritton's approach \cite{tritton} uses only the first two terms in
the right-hand side of (\ref{tempe}). It will be shown soon that two
additional trigonometric functions are needed to take into account
an influence of buoyancy near the temperature of maximum density.
The second term in the left-hand side of (\ref{teom}) generates
sines from cosines, and {\it vice versa}. In contrast to
(\ref{thetex}), therefore, the two sine functions stand in
(\ref{tempe}). Note also that our notation slightly differs from
that was used in section 17.3 of Ref. \cite{tritton}. The expansion
(\ref{tempe}) can be interpreted as a very simple example of
Galerkin methods widely used in computational fluid dynamics
\cite{fletch}. Substituting for $T-T_{0}$ the usual ansatz
$T_{1}\cos\varphi+T_{2}\sin\varphi$, the integral in the right-hand
side of (\ref{rens1}) vanishes. Indeed, one easily obtain
\begin{equation}
\int_{0}^{2\pi}(T_{1}\cos\varphi+T_{2}\sin\varphi)^{2}\sin\varphi\,\xdif\varphi
=0
\, , \label{tt12}
\end{equation}
It is for this reason that we put two additional terms into the
right-hand side of (\ref{tempe}). Elementary calculations give two
nonzero contributions, so that we reduce (\ref{rens1}) to the form
\begin{equation}
\frac{\xdif{q}}{\xdif{t}}=-\!\;\Gamma{q}+
\frac{g\varkappa}{2}\,(T_{1}T_{4}-T_{2}T_{3})
\, . \label{rens11}
\end{equation}
This equation reflects a balance of the momentum.

Following Tritton's formulation \cite{tritton}, the temperature
evolves as
\begin{equation}
\frac{\partial{T}}{\partial{t}}+\frac{q}{a}\>\frac{\partial{T}}{\partial\varphi}
=K(\varTheta_{E}-T)
\, . \label{teom}
\end{equation}
Here, the right-hand side supposes that heat is transferred through
the walls at a rate proportional to the local difference between the
external temperature and the average internal temperature.
Substituting (\ref{thetex}) and (\ref{tempe}) into (\ref{teom}) gives
\begin{align}
&\frac{\xdif{T}_{1}}{\xdif{t}}\,\cos\varphi+\frac{\xdif{T}_{2}}{\xdif{t}}\,\sin\varphi+
\frac{\xdif{T}_{3}}{\xdif{t}}\,\cos2\varphi+\frac{\xdif{T}_{4}}{\xdif{t}}\,\sin2\varphi
+\frac{q}{a}\,
\Bigl(-\!\;T_{1}\sin\varphi+T_{2}\cos\varphi-2T_{3}\sin2\varphi+2T_{4}\cos2\varphi\Bigr)
\nonumber\\
&=K\Bigl(\varTheta_{1}\cos\varphi+\varTheta_{2}\cos2\varphi-
T_{1}\cos\varphi
-T_{2}\sin\varphi-T_{3}\cos2\varphi-T_{4}\sin2\varphi
\Bigr)
\, . \label{teom1}
\end{align}
Treating these expressions as Fourier series, one gets
\begin{align}
\frac{\xdif{T}_{1}}{\xdif{t}}&=K(\varTheta_{1}-T_{1})-\frac{qT_{2}}{a}
\ , \label{t1}\\
\frac{\xdif{T}_{2}}{\xdif{t}}&=-\!\;KT_{2}+\frac{qT_{1}}{a}
\ , \label{t2}\\
\frac{\xdif{T}_{3}}{\xdif{t}}&=K(\varTheta_{2}-T_{3})-\frac{2qT_{4}}{a}
\ , \label{t3}\\
\frac{\xdif{T}_{4}}{\xdif{t}}&=-\!\;KT_{4}+\frac{2qT_{3}}{a}
\ . \label{t4}
\end{align}
It is convenient to introduce new variables
$T_{11}=\varTheta_{1}-T_{1}$ and $T_{23}=\varTheta_{2}-T_{3}$, so
that
\begin{align}
\frac{\xdif{T}_{2}}{\xdif{t}}&=-\!\;KT_{2}+\frac{q\varTheta_{1}}{a}-\frac{qT_{11}}{a}
\ , \label{t22}\\
\frac{\xdif{T}_{11}}{\xdif{t}}&=-\!\;KT_{11}+\frac{qT_{2}}{a}
\ , \label{t11}\\
\frac{\xdif{T}_{4}}{\xdif{t}}&=-\!\;KT_{4}+\frac{2q\varTheta_{2}}{a}-\frac{2qT_{23}}{a}
\ , \label{t44}\\
\frac{\xdif{T}_{23}}{\xdif{t}}&=-\!\;KT_{23}+\frac{2qT_{4}}{a}
\ . \label{t23}
\end{align}
Using $T_{11}$ and $T_{23}$ allows one to rewrite the momentum
equation (\ref{rens11}) as
\begin{equation}
\frac{\xdif{q}}{\xdif{t}}=-\!\;\Gamma{q}+
\frac{g\varkappa}{2}\,(\varTheta_{1}T_{4}-\varTheta_{2}T_{2}+T_{2}T_{23}-T_{11}T_{4})
\, . \label{rens111}
\end{equation}
The equations (\ref{t22})--(\ref{rens111}) present the basis of the
novel Lorenz-type model of thermal convection near the temperature
of maximum density.

\section{Dimensionless equations and control parameters}\label{sec3}

It is commonly accepted to reformulate equations of the above type
in dimensionless units. Together with this reformulation, proper
control parameters of the new model will be extracted here. Some
inspection allows one to accomplish a natural way to put
dimensionless units. We refrain from presenting some details here
and provide only final expressions. Let us introduce dimensionless
time $\tau=Kt$ and dimensionless functions, so that
\begin{align}
X&=\frac{q}{aK}
\, , \label{xv}\\
Y&=\sqrt{\frac{g\varkappa}{2a\Gamma{K}}}\>T_{2}
\, , \label{yv}\\
Z&=\sqrt{\frac{g\varkappa}{2a\Gamma{K}}}\>T_{11}
\, , \label{zv}\\
U&=\sqrt{\frac{g\varkappa}{2a\Gamma{K}}}\>T_{4}
\, , \label{uv}\\
V&=\sqrt{\frac{g\varkappa}{2a\Gamma{K}}}\>T_{23}
\, , \label{vv}
\end{align}
The first expression is the same that appears in Tritton's
formulation. Other expressions differ since the unit of temperature
should be put in other way due to zero $\alpha$. Namely, the factor
standing right before each temperature function in
(\ref{yv})--(\ref{vv}) replaces the factor
\begin{equation}
\frac{g\alpha}{2a\Gamma{K}}
\ . \label{factem}
\end{equation}
The control parameters are respectively defined as
\begin{align}
\pr&=\frac{\Gamma}{K}
\ , \label{prn}\\
\rp_{1}&=\sqrt{\frac{g\varkappa\varTheta_{1}^{2}}{2a\Gamma{K}}}
\, . \label{newr}\\
\rp_{2}&=\sqrt{\frac{g\varkappa\varTheta_{2}^{2}}{2a\Gamma{K}}}
\, . \label{newr2}
\end{align}
Within Tritton's approach \cite{tritton}, the parameter (\ref{prn})
plays the role of the Prandtl number. By definition, both the
parameters $\rp_{1}$ and $\rp_{2}$ cannot be negative, but may
vanish. Finally, the five dimensionless equations of the new model
are obtained as follows:
\begin{align}
\frac{\xdif{X}}{\xdif\tau}
&=-\!\;\pr{X}+\pr(\rp_{1}U-\rp_{2}Y+YV-ZU)
\, , \label{sy1}\\
\frac{\xdif{Y}}{\xdif\tau}
&=-\!\;Y+\rp_{1}X-XZ
\, , \label{sy2}\\
\frac{\xdif{Z}}{\xdif\tau}
&=-\!\;Z+XY
\, , \label{sy3}\\
\frac{\xdif{U}}{\xdif\tau}
&=-\!\;U+2\rp_{2}X-2XV
\, , \label{sy4}\\
\frac{\xdif{V}}{\xdif\tau}
&=-\!\;V+2XU
\, . \label{sy5}
\end{align}
The equation (\ref{sy1}) represents (\ref{rens111}), whereas the
equations (\ref{sy2})--(\ref{sy5}) represent
(\ref{t22})--(\ref{t23}), respectively. In contrast to generalized
Lorenz models used in Refs. \cite{curry,shen14,shen19}, the first equation
of the new system is genuinely nonlinear. This fact follows from the
equation of state (\ref{rhokp}) chosen to focus on vicinity of the
maximum density. Thus, the key distinction of the proposed model
consists of nonlinearity with respect to temperature functions in
the equation that reflects a momentum balance. The additional
equations (\ref{sy4}) and (\ref{sy5}) are very similar to
(\ref{sy2}) and (\ref{sy3}), respectively. The only distinctions are
the factor $2$ and the term $2\rp_{2}X$ in the right-hand side of
(\ref{sy4}) with the control parameter $\rp_{2}$. It will be seen
that the equations (\ref{sy2}) and (\ref{sy3}) respectively
correspond to the second and third equations appeared in Tritton's
formulation. In this sense, we developed its version suitable near
the temperature of maximum density. The analysis of the critical
points will be accomplished in the next section. In essence, the
picture that occurs in the standard Lorenz model is also reproduced
with the new model.

It is instructive to recall briefly the three equations appeared in
Tritton's approach. Then the dimensionless time and $X$ are defined
just as above. The definitions (\ref{yv}) and (\ref{zv}) should be
rewritten with the factor (\ref{factem}) standing right before the
corresponding temperature function. In dimensionless variables, the
equations read as
\begin{align}
\frac{\xdif{X}}{\xdif\tau}
&=-\!\;\pr{X}+\pr{Y}
\, , \label{st1}\\
\frac{\xdif{Y}}{\xdif\tau}
&=-\!\;Y+\rp{X}-XZ
\, , \label{st2}\\
\frac{\xdif{Z}}{\xdif\tau}
&=-\!\;Z+XY
\, , \label{st3}
\end{align}
where the control parameter
\begin{equation}
\rp=\frac{g\alpha\varTheta_{1}}{2a\Gamma{K}}
\ . \label{newr0}
\end{equation}
Here, the denominator $2a\Gamma{K}$ is of acceleration dimension.
Formally, the equations (\ref{sy2}) and (\ref{sy3}) of the new
system resemble (\ref{st2}) and (\ref{st3}), respectively. Indeed,
the temperature equation (\ref{teom}) is the same as in Tritton's
approach. This equation replaces the heat flow equation commonly
used in the Boussinesq approximation. To obtain the standard Lorenz
model \cite{lorenz63} as it commonly appears, the right-hand side of
(\ref{st3}) is rewritten with $bZ$ instead of $Z$. Also, the
parameter $\pr$ should be treated as the Prandtl number.

Convective models with a truncation different from the Lorenz
truncation have already been quoted above. There are a lot of
closely related models developed for various purposes. The five-mode
model of thermohaline convection proposed by Veronis
\cite{veronis65} uses a similar truncation of modes. The Lorenz
model can be obtained within the concept of Volterra gyrostats
\cite{agspd82,atpf99,tong}. The standard Lorenz model was also
reformulated in the sense of modulated control parameters
\cite{bbcs84,lucke85,osenda}. In addition, the authors of Ref.
\cite{osenda} compared their findings with the experimental data
reported in Ref. \onlinecite{meyer}. An eight-mode model for convection in
binary mixtures was examined in Refs. \cite{cross,lucke87}. It must be
stressed that the system (\ref{sy1})--(\ref{sy5}) differs from all
the mentioned models by nonlinearity in the first equation for
dimensionless acceleration. At the same time, the new model of
convection can also be recast in these respects. Questions of such a
kind are beyond the present consideration. In this paper, we aim to
concentrate on the system (\ref{sy1})--(\ref{sy5}) and its principal
properties.

\section{Critical points}\label{sec4}

The common approach to models of the considered type is to obtain
critical points and investigate their stability. Critical points are
defined as fixed ones, e.g., as solutions to the system of algebraic
equations derived from (\ref{sy1})--(\ref{sy5}) without dependence
on time. Thus, we have
\begin{align}
-\!\;X+\rp_{1}U-\rp_{2}Y+YV-ZU&=0
\, , \label{sa1}\\
-\!\;Y+\rp_{1}X-XZ&=0
\, , \label{sa2}\\
-\!\;Z+XY&=0
\, , \label{sa3}\\
-\!\;U+2\rp_{2}X-2XV&=0
\, , \label{sa4}\\
-\!\;V+2XU&=0
\, . \label{sa5}
\end{align}
Substituting $X=0$ immediately gives $Y=Z=U=V=0$. Let us mention
briefly the physical interpretation of the zero critical point. This
solution corresponds to the fluid remaining at rest in the loop. The
fluid temperature inside the loop is constant and equal to the
external temperature $\varTheta_{E}$ at the given height. Due to
(\ref{zv}) and (\ref{vv}), the formulas $Z=0$ and $V=0$ respectively
give $T_{11}=\varTheta_{1}-T_{1}=0$ and $T_{23}=\varTheta_{2}-T_{3}=0$.
Combining the latter with $T_{2}=T_{4}=0$ really implies that
(\ref{tempe}) reduces to (\ref{thetex}). Of course, this solution
for the state at rest can directly be obtained from (\ref{teom}).

It is clear that nontrivial critical points are allowed only for
$X\neq0$. It follows from (\ref{sa2}) and (\ref{sa3}) that
\begin{equation}
Y=\frac{\rp_{1}X}{1+X^{2}}
\ , \qquad
Z=\frac{\rp_{1}X^{2}}{1+X^{2}}
\ . \label{yzdf}
\end{equation}
In a similar manner, the equations (\ref{sa4}) and (\ref{sa5}) give
\begin{equation}
U=\frac{2\rp_{2}X}{1+4X^{2}}
\ , \qquad
V=\frac{4\rp_{2}X^{2}}{1+4X^{2}}
\ . \label{uvdf}
\end{equation}
Due to (\ref{yzdf}) and (\ref{uvdf}), it holds that
\begin{equation}
YV-ZU=\frac{2\rp_{1}\rp_{2}X^{3}}{(1+X^{2})(1+4X^{2})}
\ . \nonumber
\end{equation}
Since $X\neq0$, the first equation of our system leads to
\begin{align}
&X^{-1}(1+X^{2})(1+4X^{2})\>\bigl(X-\rp_{1}U+\rp_{2}Y-YV+ZU\bigr)
\nonumber\\
&=(1+X^{2})(1+4X^{2})-2\rp_{1}\rp_{2}(1+X^{2})+\rp_{1}\rp_{2}(1+4X^{2})
-2\rp_{1}\rp_{2}X^{2}=0
\, . \label{fsq}
\end{align}
Thus, we obtain the quadratic equation
\begin{equation}
4\xi^{2}+5\xi+1-\rp_{1}\rp_{2}=0
\, , \label{fsx}
\end{equation}
which should be solved for $\xi=X^{2}$. To exclude imaginary and
zero values for $X$, we require $\xi>0$. By a little algebra, one
has
\begin{equation}
\xi_{\pm}=\frac{-\!\;5\pm\sqrt{16\rp_{1}\rp_{2}+9}}{8}
\ . \label{xisol}
\end{equation}
Nontrivial critical points exist under the condition
\begin{equation}
\rp_{1}\rp_{2}>1
\, . \label{uncon}
\end{equation}
It is interesting that both the control parameters should be
nonzero. Moreover, the obtained constraint is a reminiscent of
complementarity relation between the position and momentum spreads
due to Heisenberg's uncertainty principle. Let one of the two
control parameters decreases. To ensure the existence of nontrivial
critical points, other parameter should increase properly. In terms
of initial quantities, the condition (\ref{uncon}) reads as
\begin{equation}
g\varkappa\varTheta_{1}\varTheta_{2}>2a\Gamma{K}
\, . \label{uncon1}
\end{equation}
Critical points of the standard model (\ref{st1})--(\ref{st3}) are
obtained as follows. Equating the right-hand sides to zero, one gets
the system of three algebraic equations. The zero critical point
$X=Y=Z=0$ exists for all values of $\rp$. For $\rp>1$, there are the
two additional solutions $X=Y=\pm\!\;\sqrt{\rp-1}$ with $Z=\rp-1$.
The fact that $X$ and $Y$ have the same sign implies that in each
case hot fluid is rising and cold fluid falling \cite{tritton}. In
terms of initial quantities, the condition $\rp>1$ reads as
\begin{equation}
g\alpha\varTheta_{1}>2a\Gamma{K}
\, . \label{unson1}
\end{equation}
The condition (\ref{uncon1}) obtained for the new model is similar
to the condition (\ref{unson1}).

Of course, analytic expressions for the critical points of our
five-dimensional system are more complicated. As a result of
calculations due to (\ref{yzdf}) and (\ref{uvdf}), we have the two
points
$\bigl(\pm\!\;X_{c},\pm\!\;Y_{c},Z_{c},\pm\!\;U_{c},V_{c}\bigr)$,
where
\begin{equation}
X_{c}=\sqrt{\xi_{+}}
\, , \qquad
Y_{c}=\frac{\rp_{1}\sqrt{\xi_{+}}}{1+\xi_{+}}
\ , \qquad
Z_{c}=\frac{\rp_{1}\xi_{+}}{1+\xi_{+}}
\ , \qquad
U_{c}=\frac{2\rp_{2}\sqrt{\xi_{+}}}{1+4\xi_{+}}
\ , \qquad
V_{c}=\frac{4\rp_{2}\xi_{+}}{1+4\xi_{+}}
\ . \label{two}
\end{equation}
These solutions correspond to the fluid circulating in the loop at a
constant speed and with a constant temperature distribution. The
different signs in the first mode representing average velocity
correspond to two possible directions of circulation, anticlockwise
and clockwise. Like nontrivial steady-state solutions of the
standard model, the sign of the second and forth modes is the same
as the sign of the first. And as before, the third and fifth modes
multiplied by cosine functions in (\ref{tempe}) are always positive.

Let us examine stability of the solution without motion. Suppose
that all the variables are small and proportional to
$\exp(\lambda\tau)$. Substituting this into the system
(\ref{sy1})--(\ref{sy5}) and neglecting nonlinear terms, we obtain
\begin{equation}
\begin{vmatrix}
\,{}-(\pr+\lambda) & -\!\;\rp_{2}\pr & 0 & \rp_{1}\pr & 0 \\
\rp_{1} & -\!\;(1+\lambda) & 0 & 0 & 0 \\
0 & 0 & -\!\;(1+\lambda) & 0 & 0 \\
2\rp_{2} & 0 & 0 & -\!\;(1+\lambda) & 0 \\
0 & 0 & 0 & 0 & -\!\;(1+\lambda)\,
\end{vmatrix}
=(1+\lambda)^{2}
\begin{vmatrix}
\,{}-(\pr+\lambda) & -\!\;\rp_{2}\pr & \rp_{1}\pr \\
\rp_{1} & -\!\;(1+\lambda) & 0 \\
2\rp_{2} & 0 & -\!\;(1+\lambda)\, \\
\end{vmatrix}
=0
\, . \label{chareq}
\end{equation}
The latter is the characteristic equation with the sign minus.
Calculating the determinant of $3\times3$ matrix leads to the
characteristic equation
\begin{equation}
(1+\lambda)^{3}\left[\lambda^{2}+(\pr+1)\lambda-\pr\rp_{1}\rp_{2}+\pr\right]
=(\lambda+1)^{3}(\lambda-\lambda_{+})(\lambda-\lambda_{-})=0
\, , \label{det33}
\end{equation}
where
\begin{equation}
\lambda_{\pm}=\frac{-\!\;(\pr+1)\pm\sqrt{(\pr+1)^{2}+4\pr(\rp_{1}\rp_{2}-1)}}{2}
\ . \label{xasol}
\end{equation}
The eigenvalue $\lambda=-\!\;1$ has multiplicity three. Only
$\lambda_{+}$ can have strictly positive real part, and this happens
under the condition (\ref{uncon}). Thus, the zero critical point
becomes unstable simultaneously with the appearance of nontrivial
critical points. Both these points represent steady solutions with
constant flow in the loop. Thus, the situation that occurs in the
standard Lorenz model is also reproduced here. As the product of
control parameters increases properly, the state without motion
loses stability so that a convection is being.

Stability of nontrivial critical points is more complicated to
analyze explicitly. To avoid diving into calculations here, main
details are presented in Appendix \ref{appa}. It will be more
instructive to discuss the physical meaning of an irregular
behavior, when the steady solutions become unstable. This
instability has a different character from the instability of the
zero point that represents fluid in the rest. In the absence of any
stable steady solutions, some kind of unsteady motion must occur
anyway. It will be exemplified in the next section that oscillations
of increasing amplitude take place. Growing pulsations of the flow
are superimposed on the mean circulation. At once, observed
trajectories still lie in a finite domain of the phase space. This
property indeed holds thought it does not follow from numerical
results alone. This regime shows seemingly random ``transitions''
between neighborhoods of the two nontrivial critical points. From a
physical viewpoint, this means a sudden change in the direction of
fluid circulation in the loop. We see again that a typical situation
occurring in the Lorenz system is resembled. The results of this
section show that the proposed model is physically meaningful and
consistent.

\section{Numerical results}\label{sec5}

This section presents the numerical analysis results of the proposed
system (\ref{sy1})--(\ref{sy5}). Of course, emphasis should be given
here to chaotic dynamics in the case of unstable critical points.
The calculations were also performed for cases when both steady
solutions were stable. They completely supported the predictions
based on linear stability analysis. If only the zero critical point
is unstable, the perturbations grow monotonically until a point in
the phase space is attracted by one of the nontrivial critical
points. In the following, we visualize the trajectories between the
neighborhoods of two nontrivial critical points. Since five modes
are involved, the points in the phase space include five
coordinates. Hence, the picture cannot be shown fully, even in three
dimensions. The role of the visualization of phase space for
understanding bifurcations such as those occurring in the Lorenz
system was emphasized in Ref. \cite{stewart}. Following the original
paper of Lorenz \cite{lorenz63}, we shall characterize dynamics by
showing projections on different planes.

\begin{figure*}
\includegraphics[height=6.8cm]{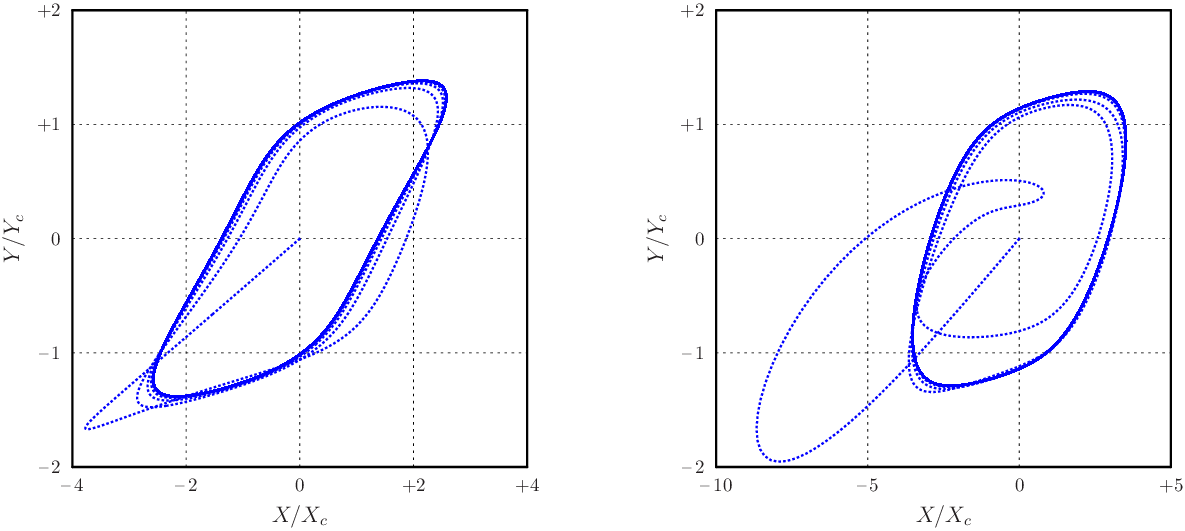}
\caption{\label{fig1} Projections on the $X-Y$ plane in phase space
of numerical solution for $\pr=5$ with $\rp_{1}=2,\ \rp_{2}=3$ on
the left and $\rp_{1}=4,\ \rp_{2}=6$ on the right. Both the cases
include $10^{5}$ iterations with the value $\delta\tau=0.01$ for
dimensionless time increment.}
\end{figure*}

\begin{figure*}
\includegraphics[height=6.8cm]{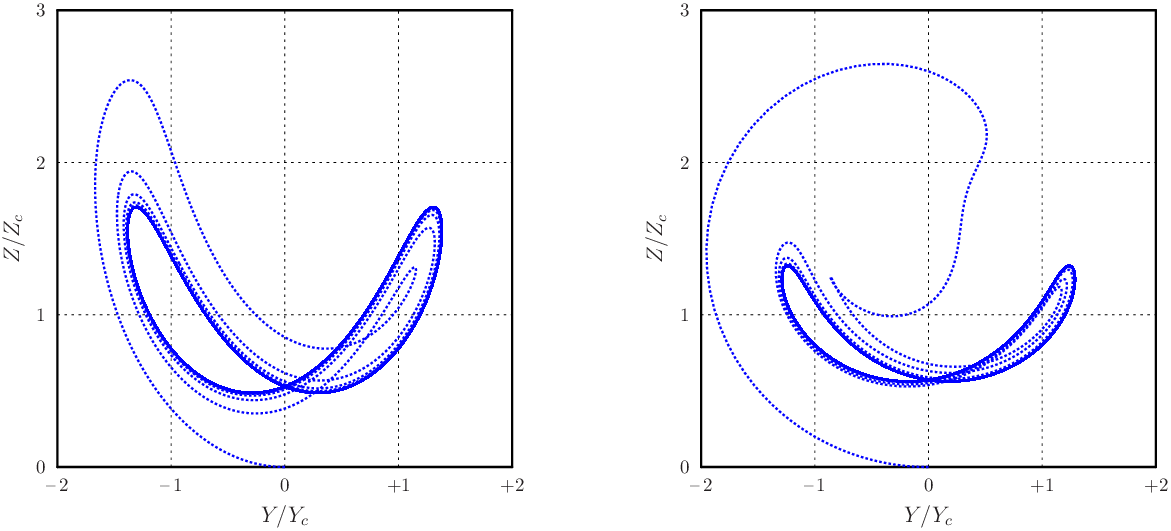}
\caption{\label{fig2} Projections on the $Y-Z$ plane in phase space
of numerical solution for $\pr=5$ with $\rp_{1}=2,\ \rp_{2}=3$ on
the left and $\rp_{1}=4,\ \rp_{2}=6$ on the right. Both the cases
include $10^{5}$ iterations with the value $\delta\tau=0.01$ for
dimensionless time increment.}
\end{figure*}

\begin{figure*}
\includegraphics[height=6.8cm]{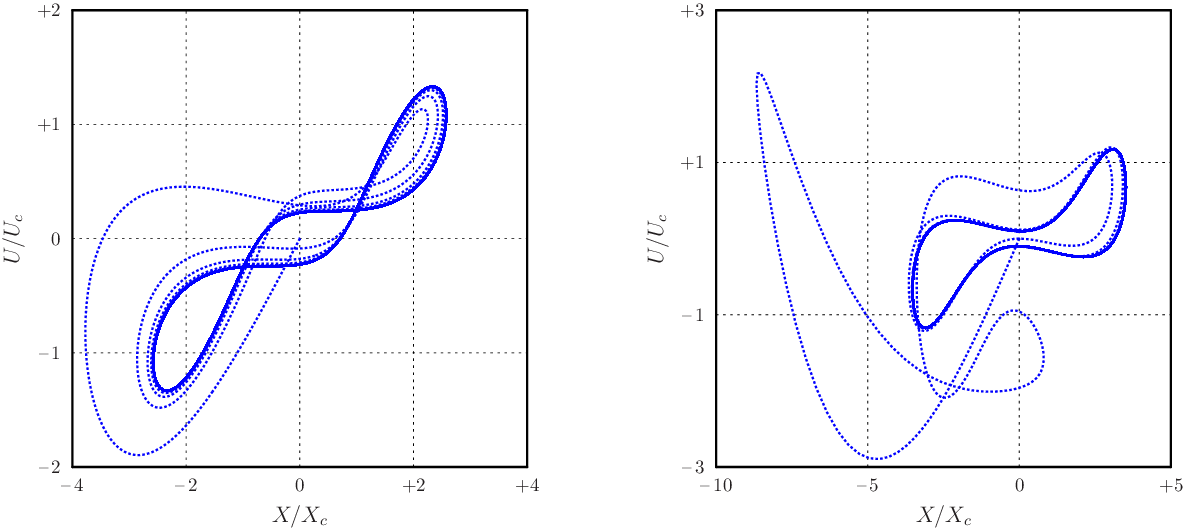}
\caption{\label{fig3} Projections on the $X-U$ plane in phase space
of numerical solution for $\pr=5$ with $\rp_{1}=2,\ \rp_{2}=3$ on
the left and $\rp_{1}=4,\ \rp_{2}=6$ on the right. Both the cases
include $10^{5}$ iterations with the value $\delta\tau=0.01$ for
dimensionless time increment.}
\end{figure*}

\begin{figure*}
\includegraphics[height=6.8cm]{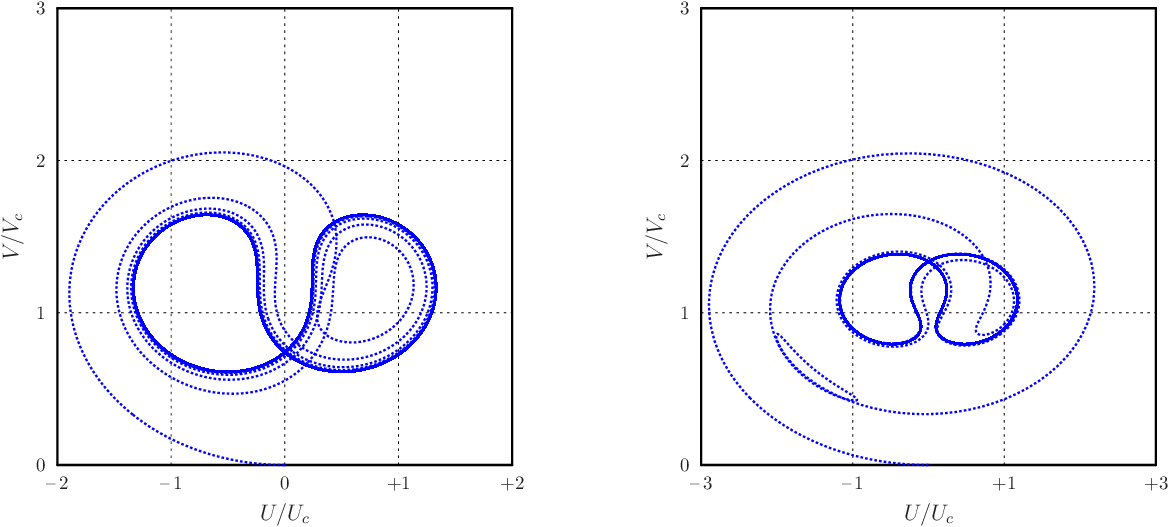}
\caption{\label{fig4} Projections on the $U-V$ plane in phase space
of numerical solution for $\pr=5$ with $\rp_{1}=2,\ \rp_{2}=3$ on
the left and $\rp_{1}=4,\ \rp_{2}=6$ on the right. Both the cases
include $10^{5}$ iterations with the value $\delta\tau=0.01$ for
dimensionless time increment.}
\end{figure*}

The equations are integrated forward in time with the use of the
classic fourth-order Runge--Kutta formula (see, e.g., section 17.1
of Ref. \cite{press07}). We refrain from presenting the well-known
content of this scheme. To obtain the results, one uses $10^{5}$
iterations of the fourth-order Runge--Kutta scheme with the value
$\delta\tau=0.01$ for dimensionless time increment. Thus, the total
interval contains $10^{3}$ dimensionless time units. To check the
stability of the numerical integration, calculations were tested
with various numbers of steps and increments. Examples of
projections on different planes are presented in Figs.
\ref{fig1}--\ref{fig4} for the following two choices of the control
parameters. The four projections are shown for $\rp_{1}=2,\
\rp_{2}=3$ on the left and for $\rp_{1}=4,\ \rp_{2}=6$ on the right,
with $\pr=5$ in both cases. Thus, the product $\rp_{1}\rp_{2}$ is
equal to $6$ in the first case and $24$ in the second case. For
$\rp_{1}=2$ and $\rp_{2}=3$ with $\pr=5$, the characteristic
equation (\ref{chareq1}) has the two roots with strictly positive
real part, viz. $\lambda=0.3978605354\pm\itt1.9524609979$ up to $10$
digits. In the second case, it also has the two roots
$\lambda=1.3160523219\pm\itt2.9811082042$. Thus, the increment is
larger in the second case.

The abscissa and ordinate of Figs. \ref{fig1}--\ref{fig4} represent
the taken variables rescaled with their values corresponding to the
positive critical point. It can be expected that observed dynamics
will become quite complicated. Under the chosen circumstances,
projections on the $X-Y$ plane are shown in Fig. \ref{fig1}. The
instantaneous state of the system is visualized by point moving
along a corresponding curve. It can be seen that the picture at the
initial stage is more complicated in the second case, with a larger
value of product $\rp_{1}\rp_{2}$ and a larger increment. After a
certain time, the clearly observed orbit attracts the current-state
point of the system. The projections on the $Y-Z$ plane are shown in
Fig. \ref{fig2}. Projections on the $X-U$ and $U-V$ planes are
plotted in Fig. \ref{fig3} and Fig. \ref{fig4}, respectively. Again,
the initial stage is more complicated in the second case with a
larger increment. At an initial time later, the current-state point
of the system is attracted by the observed orbits. In this regard,
the proposed model is similar to the standard Lorenz model and its
generalizations. Of course, further properties of this model should
also be addressed.

The word ``butterfly'' is widely used in connection with the Lorenz
system \cite{hilborn}. It is interesting that some projections in
Figs. \ref{fig2}--\ref{fig4} can also be treated to resemble a
butterfly. In this sense, the new five-dimensional model
(\ref{sy1})--(\ref{sy5}) shares some typical features with famous
examples of nonlinear dynamics. It appears that combining
theoretical analysis with extensive numerical efforts is required.
The above results were intended to motivate future studies of the
system (\ref{sy1})--(\ref{sy5}). In contrast to other modifications
of the Lorenz model, the novel model has nonlinear terms in its
first equation that follows from the Navier--Stokes equation. This
illustrates not only what may happen in the vicinity of the point of
maximum density. The proposed model can be used as a template to
build similar models in cases of more practical interest, such as
thermal convection in water layers near the maximum density. The
above numerical results demonstrate that we have rights to expect
properties similarly to those of the standard Lorenz model and its
extensions. It provides a new direction for the development of
models of the Lorenz type. It would also be interesting to visualize
three-dimensional projections of the proposed five-dimensional
model.

\section{Conclusions}\label{sec6}

We have presented a novel model of thermally induced convection in
the fluid inside the circle loop, similar to Tritton's analysis. In
contrast to the previous considerations, the fluid is assumed to be
near the point of maximum density. This topic is interesting from
its own perspective and for applications in which the temperature of
maximum density is important. In Ref. \cite{sonnino}, the experimental
data on the free convection of water near its maximum density were
compared with the theoretical predictions. The observed features
were well reproduced by the theory presented in Ref. \cite{depaz}. The
above studies consider anomalous convection in a hollow cylinder
with conducting lateral walls. In contrast, this work deals with the
convection of fluid in the loop. A buoyancy term with the square of
the excess temperature yields a five-dimensional analog of the famous
Lorenz model. Unlike some known modifications of the standard Lorenz
model, the proposed model has nonlinearity in the first equation.
The initial interest of this author to the theme was based on
temperature observations conducted in the southern basin of Lake
Baikal \cite{aab09,blprr}.

The obtained results are summarized as follows. The picture that
occurs in the standard Lorenz model is, in essence, resembled. The
nontrivial critical points of the system appear when the zero
critical point becomes unstable. The critical value depends on the
characteristics of the external temperature profile combined with
the coefficient before the square of the excess temperature in the
buoyancy term. The different signs at nontrivial critical points
correspond to two possible directions of fluid circulation. The
following facts were shown numerically. Projections on different
planes in phase space are similar to those in the standard Lorenz
model and its generalizations. Our findings support further studies
of convective models with a nonlinear buoyancy term. They deserve
more attention than they have already received. The presented model
may have potential applications in the problem of free convection in
fluid near its maximum density, when a linear relationship between
density and temperature is not adequate. This was emphasized in the
nonlinear form (\ref{rhokp}) of the thermodynamic equation of state.

\appendix

\section{On stability of nontrivial points}\label{appa}

For the two nontrivial critical points, one has the characteristic
equation with the minus sign,
\begin{equation}
\begin{vmatrix}
-\!\;(\pr+\lambda) & -\!\;\rp_{2}\pr+\pr{V}_{c} & -\!\;\pr\veps{U}_{c} & \rp_{1}\pr-\pr{Z}_{c} & \pr\veps{Y}_{c} \\
\rp_{1}-Z_{c} & -\!\;(1+\lambda) & -\!\;\veps{X}_{c} & 0 & 0 \\
\veps{Y}_{c} & \veps{X}_{c} & -\!\;(1+\lambda) & 0 & 0 \\
\,2\rp_{2}-2V_{c} & 0 & 0 & -\!\;(1+\lambda) & -\!\;2\veps{X}_{c} \\
2\veps{U}_{c} & 0 & 0 & 2\veps{X}_{c} & -\!\;(1+\lambda)\,
\end{vmatrix}
=0
\, , \label{chareq1}
\end{equation}
where $\veps=\pm\!\;1$. In general, the resulting quintic equation
is difficult to solve analytically. For a concrete choice of the
control parameters, one can calculate numerically eigenvalues of the
matrix
\begin{equation}
\begin{pmatrix}
-\!\;\pr & -\!\;\rp_{2}\pr+\pr{V}_{c} & -\!\;\pr\veps{U}_{c} & \rp_{1}\pr-\pr{Z}_{c} & \pr\veps{Y}_{c} \\
\rp_{1}-Z_{c} & -\!\;1 & -\!\;\veps{X}_{c} & 0 & 0 \\
\veps{Y}_{c} & \veps{X}_{c} & -\!\;1 & 0 & 0 \\
2\rp_{2}-2V_{c} & 0 & 0 & -\!\;1 & -\!\;2\veps{X}_{c} \\
2\veps{U}_{c} & 0 & 0 & 2\veps{X}_{c} & -\!\;1
\end{pmatrix}
 . \label{matr1}
\end{equation}
Linear stability of each of the two nontrivial critical points is
provided, when no eigenvalues have a strictly positive real part. In
fact, the Routh--Hurwitz stability criterion can also be used here.

Let us show analytically that the two nontrivial critical points
have the same eigenvalues. This physically obvious claim will serve
as an additional check of the calculations. Let us consider a
$2\times2$ block matrix
\begin{equation}
\begin{pmatrix}
\mA & \mB \\
\mC & \mD
\end{pmatrix}
 . \nonumber
\end{equation}
If $\mD$ is invertible then (see, e.g., exercise 5.30, item (b), in
the book \cite{abadir})
\begin{equation}
\ddt
\begin{pmatrix}
\mA & \mB \\
\mC & \mD
\end{pmatrix}
=\ddt(\mD)\,\ddt\bigl(\mA-\mB\mD^{-1}\mC\bigr)
\, . \label{dfor}
\end{equation}
We aim to use (\ref{dfor}) with the following blocks:
\begin{align}
\mA&=
\begin{pmatrix}
-\!\;(\pr+\lambda) & -\!\;\rp_{2}\pr+\pr{V}_{c} & -\!\;\pr\veps{U}_{c} \\
\rp_{1}-Z_{c} & -\!\;(1+\lambda) & -\!\;\veps{X}_{c} \\
\veps{Y}_{c} & \veps{X}_{c} & -\!\;(1+\lambda)
\end{pmatrix}
 , \qquad
&\mB=
\begin{pmatrix}
\rp_{1}\pr-\pr{Z}_{c} & \pr\veps{Y}_{c} \\
 0 & 0 \\
 0 & 0
\end{pmatrix}
 , \nonumber\\
\mC&=
\begin{pmatrix}
2\rp_{2}-2V_{c} & 0 & 0 \\
2\veps{U}_{c} & 0 & 0
\end{pmatrix}
 , \qquad
&\mD=
\begin{pmatrix}
-\!\;(1+\lambda) & -\!\;2\veps{X}_{c} \\
2\veps{X}_{c} & -\!\;(1+\lambda)\,
\end{pmatrix}
 . \nonumber
\end{align}
The determinant $\ddt(\mD)=(1+\lambda)^{2}+4\xi_{+}=0$ is the same
for both the points due to $\veps^{2}=1$. If $\ddt(\mD)$ vanishes
then it does this for both the nontrivial critical points
simultaneously.

Suppose further that $(1+\lambda)^{2}+4\xi_{+}\neq0$. Then the
inverse matrix exists,
\begin{equation}
\mD^{-1}=\bigl((1+\lambda)^{2}+4\xi_{+}\bigr)^{-1}
\begin{pmatrix}
-\!\;(1+\lambda) & 2\veps{X}_{c} \\
-\!\;2\veps{X}_{c} & -\!\;(1+\lambda)\,
\end{pmatrix}
 . \nonumber
\end{equation}
Elementary matrix calculations lead to the equation
\begin{equation}
\ddt\bigl(\mA-\mB\mD^{-1}\mC\bigr)=
\begin{vmatrix}
\Omega & -\!\;\rp_{2}\pr+\pr{V}_{c} & -\!\;\pr\veps{U}_{c}  \\
\,\rp_{1}-Z_{c} & -\!\;(1+\lambda) & -\!\;\veps{X}_{c} \\
\veps{Y}_{c} & \veps{X}_{c} & -\!\;(1+\lambda)\,
\end{vmatrix}
=0
\, , \label{chareq11}
\end{equation}
where
\begin{equation}
\Omega=\frac{2\pr\rp_{1}\rp_{2}(1+\lambda+2\xi_{+}\lambda)}{(1+\lambda)^{2}+4\xi_{+}}-\pr-\lambda
\, . \label{}
\end{equation}
It is clear that the resulting equation (\ref{chareq11}) is the same
for both the cases $\veps=\pm\!\;1$.

\end{document}